\documentclass[aps,pra,epsfigure,twocolumn,showpacs]{revtex4}
\usepackage{amsmath}
\usepackage{amstext}
\usepackage{latexsym}
\usepackage{psfig}
\usepackage{graphicx}
\usepackage{amsfonts}

\newcommand{\ket}[1]{\left\vert#1\right\rangle}
\newcommand{\modul}[1]{\left\vert#1\right\vert}

\newcommand{\one}{\mbox{$1 \hspace{-1.0mm}  {\bf l}$}}

\newcommand{\pro}[2]{\left\vert#1\right\rangle\left\langle#2\right\vert}

\newcommand{\proj}[3]{\left\vert#1\right\rangle_{#2}\left\langle#3\right\vert}

\newcommand{\valmed}[1]{\left\langle#1\right\rangle}
\newcommand{\num}[1]{\hat{n}_{#1}}
\newcommand{\oodet}{$O/O\hskip0.1cm\text{PD}$}
\begin{document}

\title{Generation of entangled coherent states via cross phase modulation
in a double electromagnetically induced transparency regime}
\author{M. Paternostro}
\affiliation{School of Mathematics and Physics, The Queen's University,
Belfast BT7 1NN, United Kingdom}
\author{M. S. Kim}
\affiliation{School of Mathematics and Physics, The Queen's University,
Belfast BT7 1NN, United Kingdom}
\author{B. S. Ham}
\affiliation{Center for Quantum Coherence and Communications, Electronics and Communications Research Institute, Daejeon, 305-350, South Korea}
\date{\today}

\begin{abstract}
The generation of an entangled coherent state is one of the most
important ingredients of quantum information processing using
coherent states. Recently, numerous schemes to achieve
this task have been proposed. In order to generate travelling-wave entangled coherent states, cross phase modulation,
optimized by optical Kerr effect enhancement in a dense medium in
an electromagnetically induced transparency (EIT) regime, seems to
be very promising. In this scenario, we propose a fully quantized
model of a double-EIT scheme recently proposed [D. Petrosyan and G. Kurizki, {\sl Phys. Rev. A} {\bf 65}, 33833
(2002)]: the quantization step is performed adopting a fully
Hamiltonian approach. This allows us to write effective equations
of motion for two interacting quantum fields of light that show
how the dynamics of one field depends on the photon-number
operator of the other. The preparation of a Schr\"odinger cat state, which is a
superposition of two distinct coherent states, is briefly exposed.
This is based on non-linear interaction via double-EIT of two
light fields (initially prepared in coherent states) and on a
detection step performed using a $50:50$ beam splitter and two
photodetectors. In order to show the entanglement of an entangled coherent state, we suggest to measure the
joint quadrature variance of the field. We show that the entangled
coherent states satisfy the sufficient condition for entanglement based on
quadrature variance measurement. We also show how robust our scheme is   
against a low detection efficiency of homodyne detectors.
\end{abstract}
%\pacs{03.67.Hk, 42.50.-p, 03.67.-a, 03.65.Bz}
\maketitle

%%%%%%%%%%%%%%%%%%%%%%%%%%%%%%%%%%%INTRODUZIONE%%%%%%%%%%%%%%%%%%%%%%%%%%%
\section{Introduction}
\label{introduzione}

The generation of a Schr\"odinger cat state \cite{gatto}, which is
a superposition of two distinct coherent states, and an entangled
coherent state, which is an entanglement of the coherent states,
serves the first step towards quantum information processing using
coherent states \cite{cohe}.  Numerous schemes have been proposed
in order to generate such a kind of coherent superposition
\cite{YurkeStoler,Welsch,Solano}. Cavity quantum electrodynamics
(CQED) seems to be a promising environment to this task
\cite{Haroche} and, in order to investigate their properties with
respect to decoherence, recently a scheme to generate a mesoscopic
version of a cat state using trapped ions has been proposed
\cite{Wineland}. However, most of the suggested schemes for
quantum computation using coherent states is based on
travelling-wave fields. Yurke and Stoler's suggestion to produce a
travelling-wave cat state was far from the experimental
realization because of an extremely low efficiency and a high
absorption rate of nonlinear Kerr interaction while the suggestion
primarily depends on it.

Recently, it has been proved that the interaction of two
travelling fields of light in an atomic medium
\cite{Arimondo,HarrisEIT} is able to show giant Kerr
non-linearities by means of the so-called {\sl cross phase
modulation} \cite{SchmidtedImamoglu}. Measured values of the
$\chi^{(3)}$ parameter are up to six orders of magnitude larger
than usual \cite{Hau}. This can open the way toward the use of
this kind of non-linear processes even for the very low
photon-number case \cite{HarrisHau}. Usually, the approach to such
processes is restricted to a semiclassical level: the medium is
treated quantum mechanically while the interacting fields are
assumed to be classical objects. Nevertheless, a fully quantum
treatment of non-linear dynamics is relevant with respect to many
aspects of quantum information processing. For example, huge Kerr
non-linearities can be exploited in order to perform computation,
as said above and as described in
refs.~\cite{opticalFredkingate,cohe}, to perform quantum
teleportation of an unknown state \cite{VitaliFortunatoTombesi} or
for quantum non-demolition measurements \cite{SandersMilburnQND}.
In all these examples, a complete quantum treatment of the fields
involved is required.

A full quantum analysis of the cross phase modulation problem has
been explicitly performed by Lukin and Imamoglu in
ref.~\cite{LukinImamoglu}, where a rather involved atomic system,
realized by mixing two different isotopes of the same alkali
species, has been used. In order to suggest a more feasible
experimental realization of the process, Petrosyan and Kurizki
suggested a modification of the atomic model that allows the use
of just a single species \cite{PetrosyanKurizki}. Their analysis,
however, was again semiclassical.

In this paper, we investigate the fully quantum-mechanical
description of \cite{PetrosyanKurizki} adopting a completely
Hamiltonian approach~\cite{Kryzhanovsky}. To the best of our
knowledge, this method has never been used in this context. 
We envisage in a Y$_{2}$SiO$_{5}$ crystal doped with Pr$^{3+}$ ions a 
good candidate to physically embody the atomic model we discuss: the scheme of the atomic energy levels, in this system, seems quite appropriate 
to be used for our purposes. The Pr$^{3+}$ doped Y$_{2}$SiO$_{5}$ has been used for the experimental demonstrations of EIT~\cite{ham1} and giant Kerr nonlinearity~\cite{ham2,turukhin}. Using realistic values for the atomic parameters relative to this solid state 
system we find that a giant rate of non-linearity is obtained in our fully quantum-mechanical model. We derive the relative equatios of motion 
for the involved quantum fields. This allows us to write the interaction
Hamiltonian in a form that explicitly depends on the photon-number
operators of the two quantum fields. Starting from this point, we
show how entangled coherent states and Schr\"odinger cat states
are generated when the initial states of the fields are two
independent coherent states. 

The paper is structured as follows: in Section~\ref{ilmetodo} we
describe the Hamiltonian approach we have chosen and apply it to
model cross phase modulation via
electromagnetically induced transparency (EIT)~\cite{SchmidtedImamoglu}. In
Section~\ref{doubleEIT} we apply this method to the atomic scheme
for double-EIT suggested in~\cite{PetrosyanKurizki} and we derive
the equations of motion for the quantized fields.
Section~\ref{catstates} is devoted to the generation of
entangled coherent states and Schr\"odinger cat states of light. Finally, in Section~\ref{misuradeigatti}, we describe in full detail a scheme for the detection of the entanglement in the generated entangled coherent state. The detection scheme is based on the total variance criterion for continuous variable states~\cite{duan}. 

%%%%%%%%%%%%%%%%%%%%%%%%%%%%%%%%%%%ILMETODO%%%%%%%%%%%%%%%%%%%%%%%%%%%%%%%
\section{The Hamiltonian method}
\label{ilmetodo}

The standard method to describe the interaction of electromagnetic
fields in a resonant medium is to derive the Bloch equations for
the atomic density matrix elements, which are solved in steady state
conditions. The solutions are, then, inserted into the Maxwell equations 
to show the propagation of the fields. However, when the number of fields
involved in the problem is high and the atomic system consists of
several energy levels, this procedure can be quite cumbersome.

A much simpler way to derive the field equations is given by  a
fully Hamiltonian approach \cite{Kryzhanovsky}. According to it,
the polarization of the medium can be expressed as the partial
derivative, with respect to the electric field amplitude, of the
averaged free-energy density of the atomic medium. In other words:
\begin{equation}
P=-\left\langle\frac{\partial H}{\partial E^*}\right\rangle
\end{equation}
where $H$ is the interaction part of the Hamiltonian, $E$ is the
complex amplitude of the electromagnetic field and $P$ is the
polarization of the medium \cite{Kryzhanovsky}. When several
electromagnetic fields interact with the medium, $H$ can be
expressed, following \cite{Kryzhanovsky}, as:
\begin{equation}
-H=\sum_{j}\chi^{(1)}(\omega_j)\modul{E_j}^{2}+\frac{1}{2}\sum_{ij}B_{ij}\modul{E_{i}E_j}^{2}+....
\end{equation}
where $B_{ij}$ are the diagonal elements of the  non-linear
third-order susceptibility, responsible for the non-linear terms
of the refractive index at frequency $\omega_{j}$. Thus, the
polarization due to the j-th electromagnetic field can be written
as:
\begin{equation}
\label{polarization}
\begin{split}
P_j&=\left\langle\frac{\partial H}{\partial E^{*}_j}\right\rangle{e^{-i(\omega_{j}t-k_{j}z)}}+c.c.\\
&=-\frac{Nd_{j}}{\hbar}\left\langle\frac{\partial H}{\partial \Omega^{*}_j}\right\rangle{e^{-i(\omega_{j}t-k_{j}z)}}+c.c.
\end{split}
\end{equation}
where $\Omega_{j}$ is the Rabi frequency relative to  the $j$-th
field, $d_{j}$ is the dipole matrix element of the corresponding
transition and $N$ is the density of the atomic medium.

Introducing this equation into the Maxwell-Bloch equations,  we
get rid of the atomic variables, obtaining a set of equations of
motion for the Rabi frequencies that, in the slowly varying
envelope approximation (SVEA), reads:
\begin{equation}
\left(\frac{\partial}{\partial{z}}+\frac{1}{c}\frac{\partial}{\partial{t}}\right)\Omega_{j}=-i\frac{Nd^{2}_{j}\omega_{j}}{2\hbar{\epsilon_0}c}\left\langle\frac{\partial H}{\partial \Omega^*_{j}}\right\rangle,\hskip0.3cm\forall{j}.
\end{equation}
Changing the reference frame into $\xi=z, \tau=t-z/c$, the above equation can be reduced
to:
\begin{equation}
\label{hamiltonianapproach}
\frac{\partial\Omega_{j}}{\partial{\xi}}=-i\frac{Nd^{2}_{j}\omega_{j}}{2\hbar{\epsilon_0}c}\left\langle\frac{\partial H}{\partial \Omega^*_{j}}\right\rangle.
\end{equation}

This approach has been used in ref.~\cite{Fleischhauer}  to
investigate the problem of resonant forward four-wave mixing based
on EIT. It is particularly convenient if an open-system
Hamiltonian model is used to incorporate {\it ab initio} the decay
rates of the atomic levels and if the atoms follow adiabatically
the fields evolution. While the first condition can be satisfied
using an effective complex Hamiltonian, the second point needs
more explanations.

Solving the Bloch equations that describe the atomic  density
matrix evolution, one usually invokes the so-called weak coupling
limit: the fields that couple an initially prepared, collective,
atomic state to other states of the atomic model are assumed to be
very weak (usually, there is less than one photon per atom on average). Thus,
the probability that, after the interaction, a state different
from the initial one is populated is very small. This qualifies
the initial state as a stationary state and the system will evolve
in an adiabatic fashion, following its dynamics. In these
conditions, the averaged Hamiltonian that appears in
Eq.~(\ref{hamiltonianapproach}) can be replaced by the eigenvalue
of $H$ that, in the limit of vanishing weak coupling fields, gives
the energy of the initially prepared state \cite{Fleischhauer}.
Thus:
\begin{equation}
\frac{\partial\Omega_{j}}{\partial{\xi}}=-i\frac{Nd^{2}_{j}\omega_{j}}{2\hbar{\epsilon_0}c}\left\langle\frac{\partial\lambda}{\partial\Omega^*_{j}}\right\rangle
\end{equation}
where $\lambda$ is the above cited eigenvalue of $H$.

The advantages of this approach are evident:  the knowledge of the
eigenspectrum of the single atom model suffices to derive
the field equation of motion directly. The quantization of the fields is then
performed in the canonical way, just replacing the classical field
variables in the effective Hamiltonian represented by the explicit
expression of $\lambda$ and assigning appropriate commutation
rules to them \cite{landau}. Starting from this effective, fully
quantized Hamiltonian, the quantum generalization of the equations of motion 
for the fields is easily derived. We want to stress here that
adopting this Hamiltonian approach we do not introduce any
other approximation with respect to the semiclassical case: we
just eliminate the atomic variables evolution from that of the
fields without solving the corresponding Bloch equations.

%%%%%%%%%%%%%%%%%%%%%%%%%%%%%%%%%EXAMPLE:IL MODELLODIIMAMOGLU%%%%%%%%%%%%%%%%%%%%%%%%%%%%%%%%%%%%%%%%%%%%%%%%%%%%%%%%%%%%%

Here, we propose an example to illustrate the power of the
Hamiltonian approach and to show how to get huge non linear
effects using the interaction of a field with a macroscopic atomic
ensemble in the EIT regime.

\begin{figure}[b]
\centerline{\psfig{figure=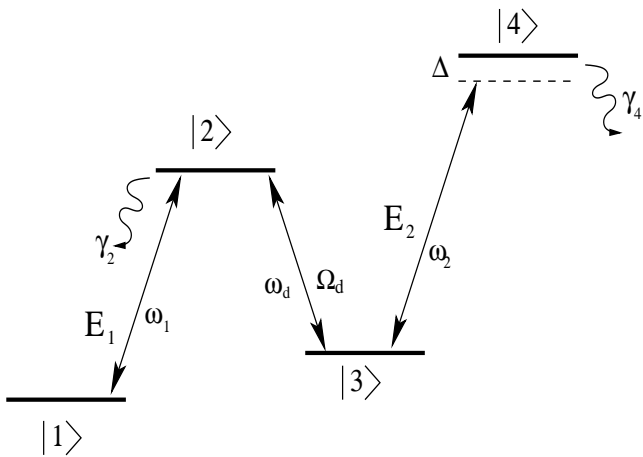,width=8.5cm,height=6.0cm}}
\caption{Sketch of the energy levels of the model by Schmidt and
Imamoglu \cite{SchmidtedImamoglu}. Fields $E_1$ and $E_2$ are
assumed to be weak with respect to the strong driving field with
fequency $\omega_d$. $\gamma_{2}$ and $\gamma_{4}$ are the decay
rates of states $\ket{2}$ and $\ket{4}$ respectively. States
$\ket{1}$ and $\ket{3}$ are assumed to be metastable. In condition
of two-photon Raman resonance, the ensemble appears transparent to
field $E_{1}$ that propagates inside it with a very slow group
velocity \cite{HarrisEIT, Arimondo}. $\Delta$ is
the detuning of the $\ket{3}\leftrightarrow\ket{4}$ transition:
this dispersive coupling induces a.c. Stark shift in the state
$\ket{3}$. This results in a shift of the refractive index curve
of the medium. Because of the steepness of this curve inside the
EIT frequency window for field $E_{1}$, the value of $\chi^{(3)}$
is strongly enhanced \cite{LukinImamoglu}.}
\label{Imamoglu}
\end{figure}

We refer explicitly to ref.~\cite{SchmidtedImamoglu}  (the atomic
model is sketched in Fig.~\ref{Imamoglu}) where, using the usual
semiclassical approach, it has been proved that giant values of
the third-order atomic susceptibility $\chi^{(3)}$ can be
obtained. This result is a consequence of the a.c. Stark shift
experienced by the assumed metastable state $\ket{3}$ because of
the dispersive coupling, induced by field $E_2$, between $\ket{3}$
and $\ket{4}$.
The Hamiltonian that describes this interaction, with a canonical transformation and introducing the decay rates of states $\ket{2}$ and $\ket{4}$,
in the basis $\left\{\ket{1}\ket{2},\ket{3},\ket{4}\right\}$ have the effective matrix representation:
%\begin{equation}
%\begin{split}
%H=\hbar\sum^{4}_{i,2}\omega_{i1}\pro{i}{i}+H_{i}
%\end{split}
%\end{equation}
%where
%\begin{equation}
%H_{i}=\hbar\{\Omega_{1}e^{-i\omega_{1}t}\pro{2}{1}+\Omega_{2}e^{-i\omega_{2}t}\pro{4}{3}+\Omega_{d}e^{-i\omega_{d}t}\pro{2}{3}+c.c.\}.
%\end{equation}
%\begin{equation}
%H_{0}=\hbar\left\{\omega_{21}\pro{2}{2}+\omega_{31}\pro{3}{3}+(\omega_{41}-\Delta)\pro{4}{4}\right\}
%\end{equation}
%$H$ can be put into the form:
%\begin{equation}
%\label{HamiltonianImamoglu}
%H'=\hbar\Delta\pro{4}{4}+\hbar\left\{\Omega_{1}\pro{2}{1}+\Omega_{2}\pro{4}{3}+\Omega_{d}\pro{2}{3}+c.c.\right\}\\
%\end{equation}
\begin{equation}
H'=\hbar
\begin{pmatrix}
0&\Omega^{*}_{1}&0&0\\
\Omega_{1}&-i\gamma_{2}&\Omega_{d}&0\\
0&\Omega^{*}_{d}&0&\Omega^{*}_{2}\\
0&0&\Omega_{2}&\Delta-i\gamma_{4}\\
\end{pmatrix}.
\end{equation}

Assuming, as in \cite{SchmidtedImamoglu},  that
$\modul{\Omega_{1}},\modul{\Omega_{2}}\ll\modul{\Omega_{d}},\Delta,\gamma_4$,
the secular equation for $H'$ results in a fourth-order polynomial
whose coefficients can be expanded in power series of
$\modul{\Omega_{1,2}}/\modul{\Omega_{d}}$. Retaining just the
first significant terms of these expansions, the relevant
eigenvalue is found to be:
\begin{equation}
\lambda_{SI}\simeq-\hbar\frac{\modul{\Omega_1}^2\modul{\Omega_2}^2}{(\Delta-i\gamma_4)\modul{\Omega_d}^2}.
\end{equation}

Taking the limit
$\modul{\Omega_1},\modul{\Omega_2}\rightarrow{0}$,  $\lambda_{SI}$
tends to zero, which is the energy of the initially prepared state
$\ket{1}$, as assumed in \cite{SchmidtedImamoglu}. 

Having $\lambda_{SI}$, the technique described in the preceding
section can be straightforwardly applied: deriving $\lambda_{SI}$,
which now represents an effective interaction Hamiltonian, with
respect to $\Omega^*_{1}$ allows us to get an expression for
$P_1$, polarization of the medium at frequency $\omega_1$.
According to Eq.~(\ref{polarization}), we have:
\begin{equation}
P_1(\omega_1)=\frac{N\modul{d_{12}}^2\modul{d_{34}}^2}
{(\Delta-i\gamma_4)\modul{\Omega_{d}}^2\hbar^3}\modul{E_2}^2E_{\omega_1}
\end{equation}
and then:
\begin{equation}
\chi^{(3)}(\omega_1)=\frac{N\modul{d_{12}}^2\modul{d_{34}}^2}
{\epsilon_{0}(\Delta-i\gamma_4)\modul{\Omega_{d}}^2\hbar^3}.
\end{equation}
This is exactly the main result obtained by Schmidt and  Imamoglu
\cite{SchmidtedImamoglu}. Taking the real part of the complex
$\chi^{(3)}$ we get the rate of non-linearity of this process.
Note that, differenly from the works in
refs.~\cite{SchmidtedImamoglu,HarrisFieldImamoglu}, here we do not
have any $\chi^{(1)}$ because of the assumed perfect resonance in
the transition $\ket{1}\leftrightarrow\ket{2}$ and the
zero atomic decay rate from state $\ket{3}$. Measured values of the non linear refractive index, 
for this model, are of the order of
$10^{-1}{cm^2/Watt}$, resulting in an enhancement of the Kerr
effect up to six orders of magnitude with respect to the best
measured values for the case of cold trapped Cs atoms \cite{Hau}.

The main result of this section has been to show that the chosen
Hamiltonian approach is able to reproduce correctly the results
obtained by solving the equations of motion for the atomic density
matrix elements. Starting from it, we will straightforwardly derive the full quantum description 
of a model for double-EIT.
%%%%%%%%%%%%%%%%%%%%%%%%%%%%%%%%%DOUBLE EIT REGIME%%%%%%%%%%%%%%%%%%%%%%%%%%%%%%%%%%%%%%%%%%%%%%%%%%%%

\section{Cross phase modulation via a double EIT effect}
\label{doubleEIT} 
\begin{figure}[b]
\centerline{\psfig{figure=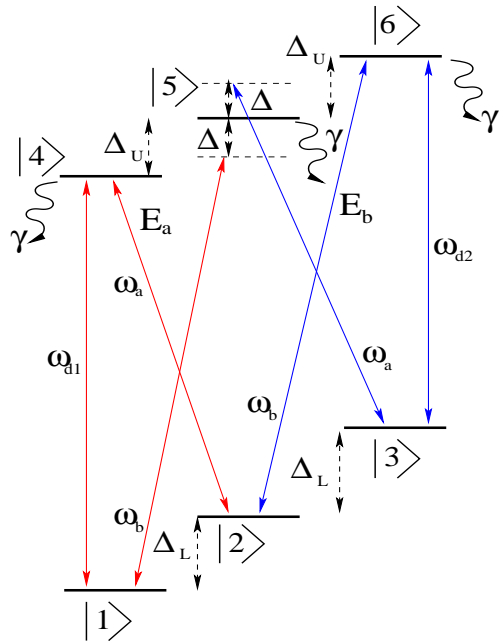,width=6.5cm,height=8.5cm}}
\caption{This figure shows the atomic model used to get a
double-EIT regime for fields $E_a$, with frequency $\omega_a$, and
$E_b$, with $\omega_b$. The fields that have frequencies
$\omega_{d1},\hskip0.1cm\omega_{d2}$ are assumed to be classical, in the
sense that their intensities is much greater than that of $E_a$
and $E_b$. The splitting $\Delta_U$ is assumed to be different
with respect to $\Delta_L$.
$\modul{\Delta}=\modul{\Delta_U-\Delta_L}$ is the detuning of
field $E_a$ relatively to the transition
$\ket{1}\leftrightarrow\ket{5}$ and of field $E_b$ with respect to
$\ket{3}\leftrightarrow\ket{5}$. The excited states decay rates
are assumed to be equal to $\gamma$, for sake of simplicity.} 
\label{Petrosyan}
\end{figure}

We refer again to Fig.~\ref{Imamoglu} for the
details of the following discussion. As explained above, in the
EIT regime, the field $E_1$ travels in the medium with a very slow
group velocity (17 $m/sec$ in \cite{Hau} and 45$m/sec$ in~\cite{turukhin}) while $E_2$ has a very high propagation velocity.
Harris and Hau proved \cite{HarrisHau} that the total phase shift
experienced by field $E_1$ is limited by the time that the
faster of the two fields spends inside the medium. The efficiency of the
non-linear interaction is, thus, strongly affected by any velocity
mismatch. In order to get rid of this bottleneck, strategies to
induce EIT for both $E_{1}$ and $E_{2}$ ({\sl double}-EIT regime)
have been developed. This will maximize the interaction time,
optimizing the efficiency of the process. While the scheme
suggested in ref.~\cite{LukinImamoglu}, even if extremely
stimulating, seems to be hard to be experimentally realized,
Petrosyan and Kurizki \cite{PetrosyanKurizki} proposed another
scheme for double-EIT to simplify the model. Even if it implies
a complication of the atomic energy spectrum, it appears simpler
under a realizable point of view. The energy scheme is sketched in
Fig.~\ref{Petrosyan}: it involves a six-level atomic configuration
and four electromagnetic fields. A magnetic field splits
metastable triplet $\left\{\ket{1},\ket{2},\ket{3}\right\}$ by
$\Delta_{L}$ and the excited triplet
$\left\{\ket{4},\ket{5},\ket{6}\right\}$ by
$\Delta_{U}\neq\Delta_{L}$. Transition
$\ket{2}\leftrightarrow\ket{5}$ is assumed to be forbidden, while
$\ket{2}$ is resonantly coupled to states $\ket{4}$ and $\ket{6}$
by means of the two very weak probes $E_a$ and $E_b$,
respectively. These two fields couple transitions
$\ket{1}\leftrightarrow\ket{5}$ and
$\ket{3}\leftrightarrow\ket{5}$ with a detuning
$\modul{\Delta}=\modul{\Delta_{U}-\Delta_{L}}$. The couplings
$\ket{1}\leftrightarrow\ket{4}$ and
$\ket{3}\leftrightarrow\ket{6}$ are realized by two classical,
intense fields of different frequencies but equal Rabi
frequencies. In these conditions, the system divides itself into
two parts. For  the subsystem composed of
$\ket{1},\ket{4},\ket{2},\ket{5}$, EIT is induced for field $E_a$
while an a.c. Stark shift effect on state $\ket{1}$  is determined
by  $E_{b}$ to generate the required non-linear interaction. For
the subsystem composed of $\ket{3},\ket{6},\ket{2},\ket{5}$, an
analogous discussion can be done interchanging $E_a$ and $E_b$.
The two subsystems are related via the non-resonant couplings
involving $\ket{5}$. The double-EIT regime is, thus, established.

As we have discussed above, a Hamiltonian approach reveals its
advantages when several atomic levels are involved. In these
cases, even if a Maxwell-Bloch approach is still possible, the
procedure itself is rather uncomfortable. Furthermore, the
generalization to a fully quantized version of a non-linear
process can be hard to perform \cite{Fleischhauer}.

For the system described in Fig.~\ref{Petrosyan}, we write the
Hamiltonian in the interaction picture:
\begin{equation}
\label{Hamiltoniana}
\begin{split}
H'&=\hbar\Delta\pro{5}{5}+\hbar\left\{\Omega_{d}\pro{4}{1}+\Omega_{d}\pro{6}{3}\right.\\
&+\Omega_{a}\pro{4}{2}+\Omega_{b}\pro{5}{1}+\Omega_{b}\pro{6}{2}\\
&+\left.\Omega_{a}e^{-2i\Delta{t}}\pro{5}{3}+c.c.\right\}.
\end{split}
\end{equation}

To show that our method is able to mimic the results obtained, at
the semiclassical level, by the approach chosen in
ref.~\cite{PetrosyanKurizki}, we appropriately change the signs in
front of each Rabi frequency in Eq.~(\ref{Hamiltoniana}) and we
introduce the excited states decay rates. In ref.~\cite{PetrosyanKurizki}, the sign in front of each Rabi frequency is chosen according to the Clebsch-Gordan coefficient of the corresponding transition. Here, this is performed in a {\sl phenomenological} way just to match our model with the one reported there. We finally get the following matrix representation of $H'$ for the atomic basis
$\left\{\ket{1},\ket{2},\ket{3},\ket{4},\ket{5},\ket{6}\right\}$:
%\begin{widetext}
\begin{equation}
\label{matriceHamiltoniana}
H'=\hbar
\begin{pmatrix}
0&0&0&\Omega^{*}_{d}&-\Omega^{*}_{b}&0\\
0&0&0&\Omega^{*}_{a}&0&-\Omega^{*}_{b}\\
0&0&0&0&\Omega^{*}_{a}e^{2i\Delta{t}}&-\Omega^{*}_{d}\\
\Omega_{d}&\Omega_{a}&0&-i\gamma&0&0\\
-\Omega_{b}&0&\Omega_{a}e^{-2i\Delta{t}}&0&\Delta-i\gamma&0\\
0&-\Omega_{b}&-\Omega_{d}&0&0&-i\gamma\\
\end{pmatrix}.
\end{equation}
%\end{widetext}

The solution of the Schr\"odinger equation for a state
\begin{equation}
\ket{\phi}=\sum^{6}_{i,1}A_{i}(t)\ket{i}
\end{equation}
%$\ket{\phi}=A_{1}(t)\ket{1}+A_{2}(t)\ket{2}+A_{3}(t)\ket{3}+A_{4}(t)\ket{4}+A_{5}(t)\ket{5}+A_{6}(t)\ket{6}$
is obtained assuming the weak field limit
$\modul{\Omega_d},\gamma,\Delta\gg\modul{\Omega_{a,b}}$ and that
both $\gamma^{-1}$ and $\Delta^{-1}$ are larger than $T$, the
characteristic interaction time of the applied fields with the
atomic medium. Under these conditions, we can use the SVEA for the
atomic probability amplitudes $A_{i}(t)$ ($i=1,..,6$) and for the
field amplitudes $E_a$ and $E_b$: this is equivalent to require
that the amplitudes of the applied weak fields do not change too
much during $T$. Neglecting the highly oscillating terms, in such
a way that a kind of rotating wave approximation (RWA) is
performed, the probability amplitudes reach stationary values. Note that this
second assumption agrees with an adiabatic solution of the
equations of motion.

If all the atoms in the ensemble are initially prepared  in state
$\ket{2}$, in the weak fields limit we can take
$A_{2}(t)\simeq{1}$, $\forall{t}\le{T}$.
Ref.~\cite{PetrosyanKurizki} shows that the atomic polarizability
of the medium at frequency $\omega_{a}$ is given by:
\begin{equation}
\label{soluzionePetrosyan}
\alpha_{a}=\frac{2i\alpha_{0}\gamma\modul{\Omega_{b}}^2}{(\gamma+i\Delta)\modul{\Omega_{d}}^2},
\end{equation}
where $\alpha_{0}=N\sigma_{0}$ and $\sigma_{0}$ is the resonant
absorption cross section, generally defined by
$\sigma_{0}=\frac{\modul{d}^{2}\omega}{2\epsilon_{0}c\hbar\gamma}$.
Eq. (\ref{soluzionePetrosyan}) shows explicitly the effect of the
cross phase modulation induced by the interaction between the two
weak (but classical) fields: the polarizability at frequency
$\omega_{a}$, due to field $E_{a}$, depends on the intensity of
field $E_{b}$. Since a completely analogous expression holds for
the polarizability $\alpha_{b}$ at frequency $\omega_b$, the cross
effect is evident. Here, we are assuming that the atomic ensemble is mantained at a sufficiently low temperature to discard any Doppler broadening. 
Rigorously speaking, the thermal distribution of the atomic velocities has an influence on the value of the susceptibility of the medium, 
that has to be averaged over the velocity distribution function. If the temperature of the sample is kept low (orders of $10^2$ $ nK$ in \cite{Hau}) 
and if we adopt a co-propagating beams configuration in order to get rid of residual Doppler shifts, the broadening can be made small and the average 
can be avoided \cite{PetrosyanKurizki}. 

Introducing $\alpha_{a}$ into the equation of motion for $E_{a}$ we get the solution:
\begin{equation}
\label{soluzioneclassica}
{E}_{a}(L,t)=E_{a}\left(0,t-\frac{L}{c}\right)\exp\left\{i\int^{L}_{o}\alpha_{a}dz\right\}
\end{equation}
where $L$ is the interaction length of the fields.  $E_{a}$
experiences, thus, a phase shift due to the presence of the second
field \cite{LukinImamoglu,PetrosyanKurizki}.

We now apply the Hamiltonian approach, to show that the results
obtained solving perturbatively both the Maxwell equations for the
fields and the equations for the atomic probability amplitudes can
be obtained just by looking for the eigen-energy of the system that,
for $\modul{\Omega_{a,b}}\rightarrow{0}$, gives the energy of
state $\ket{2}$ (that is the initial state of the system).

The secular equation for the matrix $H'$  given in
Eq.~(\ref{matriceHamiltoniana}) is a six-order polynomial
expression whose coefficients depend on the Rabi frequencies
$\Omega_{a,b,d}$. In the weak field limit, we use a series
expansion of ${\Omega_{a,b}}/{\Omega_{d}}$. Retaining just the
first orders and performing that kind of RWA that allows us to
neglect all the very highly oscillating terms, we finally get:
\begin{equation}
\label{soluzionemia}
\lambda\simeq\frac{2\hbar\modul{\Omega_{a}}^{2}\modul{\Omega_{b}}^{2}\modul{\Omega_{d}}^{2}}{i\gamma\modul{\Omega}^4-\Delta\modul{\Omega_{d}}^{2}\modul{\Omega}^{2}}
\end{equation}
with $\modul{\Omega}=\sqrt{\modul{\Omega_{a}}^{2}+\modul{\Omega_{b}}^{2}+\modul{\Omega_{d}}^{2}}$.

As pointed out, this equation has to be contrasted with that for
the eigen-energy of state $\ket{2}$ in absence of the weak probes.
If these fields are absent, it is easy to verify that the atomic
model shown in Fig.~\ref{Petrosyan} can be mapped into an effective
five-level system that does not include state $\ket{2}$. The
diagonalization of the resulting Hamiltonian (obtained from
Eq.~(\ref{matriceHamiltoniana}) getting rid of the second row and
column of the matrix) shows that the states $\ket{1}$ and
$\ket{4}$ are  dressed by the intense field with frequency
$\omega_{d1}$, while the field with frequency $\omega_{d2}$
dresses the transition $\ket{6}\leftrightarrow\ket{3}$. This shows
that $\ket{2}$ is the only state that, in absence of weak fields
but with the strong classical ones shined on the medium,  has zero
energy. Since, for $\modul{\Omega_{a,b}}\rightarrow{0}$, we have
$\lambda\rightarrow{0}$, Eq.~(\ref{soluzionemia}) is the right
solution.

Assuming once more the weak field limit, we have
$\modul{\Omega}\simeq\modul{\Omega_{d}}$ and the expression for
$\lambda$ can be approximated to:
\begin{equation}
\label{soluzionemiaapprox}
\lambda\simeq\frac{2\hbar\modul{\Omega_{a}}^2\modul{\Omega_{b}}^2}{(i\gamma-\Delta)\modul{\Omega_{d}}^2}.
\end{equation}

The partial derivative of Eq.~(\ref{soluzionemiaapprox}) with
respect to $\Omega^*_{a}$ gives us an explicit expression
for the polarization of the medium at frequency $\omega_{a}$ and
the equation of motion for $\Omega_{a}$. The latter, finally,
reads:
\begin{equation}
\frac{\partial\Omega_{a}}{\partial\xi}=\frac{2iN\sigma_{0}\gamma\modul{\Omega_{b}}^2}{(\gamma+i\Delta)\modul{\Omega_{d}}^2}\Omega_{a},
\end{equation}
that exactly corresponds to the result semi-classically obtained
in ref.~\cite{PetrosyanKurizki}. In the same way, the partial
derivative with respect to $\Omega^*_{b}$ leads to the
polarizability at frequency $\omega_{b}$ and to the equation of
motion for the Rabi frequency $\Omega_{b}$. Because of the
symmetry of the system with respect to the two fields, we easily
recognize that their group velocities are both equal to
$v_{g}\simeq{\modul{\Omega_{d}}^2}/{N\sigma_{0}\gamma}\ll{c}$. This inequality 
holds using the values reported
in \cite{SchmidtedImamoglu}. As explained above, this equally
slow propagation of the two fields inside the medium optimizes the
cross phase modulation effect.

To have an efficient non-linear process, the rate of
two-photon-absorption has to be negligible with respect to the
rate of non-linearity. Since the former quantity is proportional
to the imaginary part of the polarizability  while the latter is
proportional to his real part \cite{HarrisHau}, we can consider
the following figure of merit for the non-linear interaction:
\begin{equation}
\label{figuremerit}
\eta\equiv\frac{\Re\left\{\alpha_{a}\right\}}{\Im\left\{\alpha_{a}\right\}}=\frac{\Delta}{\gamma}.
\end{equation}

If the experimental conditions are such that  $\Delta\gg\gamma$,
then any absorption can be neglected and the process can be seen
just as a mutually induced phase-shift of the fields. Note that
this is fully consistent with the requirement advanced in the
original theory of giant Kerr non-linearity by Schmidt and
Imamoglu \cite{SchmidtedImamoglu}. Adopting the values chosen in
ref.~\cite{PetrosyanKurizki}, an interaction length of the order
of $cm$ and an interaction time of some $\mu{sec}$ lead, for two
focused beams $E_a$ and $E_{b}$, to a total phase shift (obtained
integrating $\Re\left\{\alpha_{a}\right\}$ over the interaction
length) that can easily reach $\pi$. With these orders of
magnitude, the total two-photon-absorption probability is smaller
than $1\%$.

The quantization of the fields, now, proceeds as follows: we replace the
complex Rabi frequencies that appear in
Eq.~(\ref{soluzionemiaapprox}) with the positive and negative
frequency components of the corresponding field operators (that
satisfy the bosonic commutation rules
$[\hat{\Omega}_{i},\hat{\Omega}^{\dagger}_{j}]\propto\delta_{ij}\hat{\one}$,
with $\delta_{ij}$ the Kronecker symbol, $\hat{\one}$ the identity
operator and $i,j=a,b$), multiply the expression that is thus
obtained by the density of the atoms in the ensemble ($N$) and
integrate over the interaction volume $V=AL$, with $A$ the
effective cross section of the fields. Following this recipe, we
get an effective Hamiltonian operator that describes, in a
completely quantum picture, the non-linear interaction of two
quantum fields that propagate inside a dense medium in condition
of double-EIT:
\begin{equation}
\label{Hamiltonianeffettiva}
\hat{H}_{eff}=\frac{2\hbar{AN}}{(i\gamma-\Delta)}\int^{L}_{0}
\frac{\hat{\Omega}^{\dagger}_{a}\hat{\Omega}_{a}\hat{\Omega}^{\dagger}_{b}\hat{\Omega}_{b}}
{\modul{\Omega_{d}}^2}dz.
\end{equation}
%\begin{equation}
%\label{Hamiltonianeffettiva}
%\hat{H}_{eff}=\frac{2\hbar{AN}}{(i\gamma-\Delta)}\left\int^{L}_{0}\frac{\hat{\Omega}^{\dagger}_{a}\hat{\Omega}_{a}\hat{\Omega}^{\dagger}_{b}\hat{\Omega}_{b}}{\modul{\Omega_{d}}^2}dz\right.{.}
%\end{equation}

For the case of pulses propagating inside the non-linear medium, we can follow a treatment analogous to that developed in~\cite{continuumfield}. Thus, adopting the narrow bandwidth approximation and assuming a finite range of frequencies involved in the superpositions that build up the pulses, we introduce the slowly varying positive frequency operator $\hat{\Omega}_{a}(z,t)=d_{24}\sum_{k}\sqrt{\frac{\omega^{car}_{a}}{2\hbar\epsilon_{0}V_{q}}}\hat{a}_{k}(t)e^{-i(\omega_{k}-\omega^{car}_{a})z/c}$ and the analogous for $\hat{\Omega}_{b}(z,t)$, where we explicitly introduced the annihilation operators
$\hat{a}_{k}$. Here, $V_{q}$ is the quantization volume, $k$ is a label for the different wavelengths appearing in the superposition and $\omega^{car}_{a}$ is the central ({\it carrier}) frequency of the pulse: the narrow bandwidth approximation consists in assuming that the width of the pulses, in the frequency domain, is smaller than the carrier frequency itself. When the spatial integration is carried on, assuming that the medium length is longer than all the wavevelengths in the pulses, the main contribution to $\hat{H}_{eff}$ is due to terms as $\sum_{k}\hat{a}_{k}^{\dag}(t)\hat{a}_{k}(t)$ (and the same for field $E_{b}$) which define the total photon number operator in the pulse $\hat{N}_{a}$ ($\hat{N}_{b}$). In the case of single-mode field, as a cw laser beam, the sum that appears in the definition of $\hat\Omega_{a,b}(z,t)$ collapses and $\hat{\Omega}_{a}(z,t)=d_{24}\sqrt{\frac{\omega_{a}}{2\hbar\epsilon_{0}V_{q}}}\hat{a}e^{-i(\omega_{a}{t}-k_{a}z)}$. Assuming $\eta\gg1$ and properly
collecting all the non-operatorial quantities into a rate of
non-linearity $\chi$, we can write
Eq.~(\ref{Hamiltonianeffettiva}) as:
\begin{equation}
\label{catHamiltonian}
\hat{H}_{eff}=\hbar\chi\hat{a}^{\dagger}\hat{a}\hat{b}^{\dagger}\hat{b}
\end{equation}
with
\begin{equation}
\label{chialto}
\chi=\Re\left\{\frac{N\omega_{a}\omega_{b}\modul{d_{24}}^2\modul{d_{26}}^2}{2\hbar^2\epsilon_{0}^2(i\gamma-\Delta)\modul{\Omega_{d}}^2V}\right\},
\end{equation}
where we have assumed that the interaction volume coincides with the quantization one. This is the counterpart, specialized to the particular  atomic
model we have adopted here, of the rate of non-linearities obtained by Lukin and Imamoglu in~\cite{LukinImamoglu}.
 
Let us briefly turn to some experimental details. A promising candidate to embody the atomic model we used for the double-EIT regime is a crystal of Y$_{2}$SiO$_{5}$ doped with Pr$^{3+}$ ions (Pr:YSO)~\cite{equall}, both for an interesting similarity between the energy-level scheme described here and that of the transition ${^3H}_{4}\rightarrow{^1D}_{2}$ in this crystal and for the possibility we have, in a solid state system, to limit the effect of the Doppler broadening. 
This solid state system is notable for its relatively narrow-linewidth EIT that, very recently, enabled the observation of ultraslow group velocity (approximatively $45$ m/sec) and storage of light pulses (measured delay times greater than $65$ $\mu$sec)~\cite{turukhin}. For the ${^3H}_{4}\rightarrow{^1D}_{2}$ transition at a wavelength of $\sim600$ nm considered in~\cite{turukhin}, the ground state population lifetime is of the order of minutes. The crystal sample can be taken as long as $1$ mm and the laser beams used in the non-linear interaction can be focused, by a lens, to have a typical diameter of $100$ $\mu$m (Full-Width-at-Half-Maximum). A realistic value for the excited states decay rate $\gamma$ range between $10$ and $100$ kHz (close to the measured values for a sample at a temperature of $5$ K) that, for detuning $\Delta\sim1$ MHz and coupling field Rabi frequency $\modul{\Omega_{d}}\sim1$ MHz allows us to consider $\gamma\ll\Delta,\hskip0.1cm\modul{\Omega_{d}}$, in accordance with our treatment. The electric dipole matrix elements for the system in exam are three orders of magnitude weaker than those, for the same range of frequencies, in alkali atoms. Taking, as typical values for an alkali atom $d\sim10^{-29}$ Cm, we can assume $\modul{d_{24}}\simeq\modul{d_{26}}\sim10^{-32}$ Cm. Finally, the atomic density is taken as $N\sim10^{15}$ cm$^{-3}$. Putting these values in Eq.~(\ref{chialto}) we find that interaction times $\tau$ in the range of $\mu$sec allow to reach a cross-phase shift $\varphi=\chi\tau\sim\pi$ even when the intensity of the beams involved (proportional to $\valmed{\hat{n}_{a,b}}$) is no more than a few photons. We will see in the next section how important, for the purposes of this paper, is the ability to get a $\pi$ cross-phase shift of the interacting fields.  

For the case of pulses interacting in the atomic sample, some experimental difficulties rise. First of all, effects of diffraction, focusing and defocusing on $E_{a}$ and $E_{b}$ are to be considered: they are due to the transverse intensity profile of $\Omega_{d}$ that leads to variations in the radial refractive index experienced by the weak probes (this effect is known in literature as {\it electromagnetically-induced-focusing} (EIF)~\cite{moseley}). EIF 
changes the size of the weak beams from point to point inside the medium, thus modifying the interaction volume (that becomes a function of 
the position) and influencing the rate of non-linearities. The effect, present even in the cw regime, is less controllable in the case of pulses because of the different frequencies involved in the propagating packets. These non-linear effects, indeed, seem to depend on the sign of 
the detunings between the components of the probes and the frequency of the atomic transitions that they guide. This implies, for example, focusing for some components of a probe pulse accompanied by defocusing of all the other components (of the same pulse) that have a detuning of opposite sign. The result is that different harmonics, in a pulse, are subject to different radial evolutions, thus complicating the control on the dynamics of the field itself.

Another relevant point to be treated is that of the different phase shifts acquired by the different parts of the cross-interacting pulses. As it can be seen solving the Heisenberg equations of motion for the field operators $\hat{E}_{a}$ and $\hat{E}_{b}$ interacting as described in Eq.~(\ref{Hamiltonianeffettiva}), and as it is proved in~\cite{LukinImamoglu}, the evolution of the probe fields is given by:
\begin{equation}
\label{pulsespropagation}
\hat{E}_{a,b}(L,t)=\hat{E}_{a,b}(0,t')e^{\left\{i\tilde\chi\hat{E}_{b,a}^{\dagger}(0,t')\hat{E}_{b,a}(0,t')\right\}}
\end{equation}  
with $t'=t-L/v_{g}$ and ${\chi}$ the rate of non-linearity specific for this case (that reduces to Eq.~(\ref{chialto}) when we consider cw fields). Note that this solution is the exact quantum analogous of what has been found, in the semiclassical approach, in Eq.~(\ref{soluzioneclassica}). Thus, the phase shift experienced by pulse $a$ ($b$) depends on the total number of photons in pulse $b$ ($a$) at the earlier time $t'$ and this number changes with the amplitude of $E_{b}$ ($E_{a}$). The effect is that different parts of a propagating pulse acquire different phase shifts with respect to each other. It modifies, in essence, the relative phase relations between the pulse components and distorts the pulse shape. Spectral width enlargement, for example, can be a detrimental consequence for the purely dispersive propagation inside the atomic medium: if some harmonics of the evolving pulses exit from the EIT transparency window, they will be strongly absorbed by the no-more-transparent medium. Usually, a way to bypass this kind of problem is to arrange the pulses to be within the EIT window since their entrance into the non-linear medium, properly choosing their shape and spectral width. For adiabatic evolution and for an optically thick medium, the pulses will last in the non absorptive region of the refractive index of the medium through all the interaction time~\cite{darkpolaritons}. But this result has been proved just for a simple atomic $\Lambda$ system and it needs a deeper analysis when we refer to the atomic energy scheme we describe in this paper. In our opinion, all these points have to be further investigated, even seeking for an experimental verification of their real influence, in order to have a complete comprehension of the kind of control we can reach for the cross-phase interaction of two weak fields in a medium that exhibits giant Kerr non-linearities. 

For the sake of simplicity, but without lacking the necessary experimental realism, in what follows we treat just the case of single cw mode propagation and, from now on, we proceed taking in consideration the effective Hamiltonian in Eq.~(\ref{catHamiltonian}).
 
%%%%%%%%%%%%%%%%%%%%%%%%%%%%%%%%GATTI%%%%%%%%%%%%%%%%%%%%%%%%%%%%%%%%%%%%%%%%%%%%%%%%%%%%%%%%%%%%%%%%%%%%%%%%%%%%

\section{Schr\"odinger cat states generation}
\label{catstates}

In this section we want to apply the results obtained with our
fully quantized picture of the cross phase modulations induced by
double-EIT in order to show that this specific system can be used
to produce a Schr\"odinger cat state \cite{gatto} of a single mode of field.

Given the interaction Hamiltonian~(\ref{catHamiltonian}),  we
derive the equations of motion for the annihilation operators
$\hat{a}$ and $\hat{b}$. Adopting the usual notation
$\num{a}=\hat{a}^{\dagger}\hat{a},\hskip0.1cm\num{b}=\hat{b}^{\dagger}\hat{b}$
for the photon-number operators of the two fields, these are:
\begin{equation}
\begin{split}
\partial_{t}\hat{a}&=-i\chi[\hat{a},\num{a}\num{b}]=-i\chi\hat{a}\num{b}\\
\partial_{t}\hat{b}&=-i\chi[\hat{b},\num{a}\num{b}]=-i\chi\hat{b}\num{a}
\end{split}
\end{equation}
which, arranging the time scale in such a way that $t=0$  is the
instant in which the interaction starts, lead to the
time-dependent operators:
\begin{equation}
\hat{a}_{out}(t)=e^{-i\chi{t}\num{b}}\hat{a}(0)\hskip0.8cm\hat{b}_{out}(t)=e^{-i\chi{t}\num{a}}\hat{b}(0).
\end{equation}
Note that, in the approximation of a pure cross shift process,
$\num{a,b}$ do not evolve because of the non-absorption character
of the interaction considered. Defining $\varphi=\chi{t}$, the
above evolution is attributed to the action of the unitary
time-evolution operator
$\hat{U}(\varphi)=e^{-i\varphi\num{a}\num{b}}$ on the field
operators. We want to specialize the present analysis to the case
in which the initial state of the two interacting fields is
$\ket{\psi(0)}_{ab}=\ket{\alpha}_{a}\otimes\ket{\gamma}_{b}$,
where the coherent state
$\ket{\alpha}_{a}=\hat{D}_a(\alpha)|0\rangle_a$ with the
displacement operator
$\hat{D}_a(\alpha)=\exp(\alpha\hat{a}^{\dagger}-\alpha^{*}\hat{a})$
\cite{Glauber}.  The coherent state $|\gamma\rangle_b$ has been
defined likewise.

The evolution of the initial state by means of $\hat{U}(\varphi)$ can be expressed as follows:
\begin{equation}
%\begin{split}
\label{evolvedstate}
\ket{\psi(t)}_{ab}=\tilde{\hat{D}}_{a}(\alpha)\tilde{\hat{D}}_{b}(\gamma)\ket{0}_{a}\otimes\ket{0}_{b},
%\end{split}
\end{equation}
where we have defined the time-dependent displacement operators:
\begin{equation}
\label{evolvdisp}
\left\{
\begin{aligned}
\tilde{\hat{D}}_{a}(\alpha)&\equiv\hat{U}(\varphi)
\hat{D}_{a}(\alpha)\hat{U}^{-1}(\varphi)=e^{(\alpha\tilde{\hat{a}}^{\dagger}-\alpha^{*}\tilde{\hat{a}})}
\\
\tilde{\hat{D}}_{b}(\gamma)&\equiv\hat{U}(\varphi)
\hat{D}_{b}(\gamma)\hat{U}^{-1}(\varphi)=e^{(\gamma\tilde{\hat{b}}^{\dagger}-\gamma^{*}\tilde{\hat{b}})}
\end{aligned}
\right.
\end{equation}
with
$\tilde{\hat{a}}^{\dagger}=e^{-i\varphi\num{a}\num{b}}\hat{a}^{\dag}e^{i\varphi\num{a}\num{b}}$
and
$\tilde{\hat{b}}^{\dagger}=e^{-i\varphi\num{a}\num{b}}\hat{b}^{\dag}e^{i\varphi\num{a}\num{b}}$.
Explicitly using the operator expansion theorem \cite{MandelWolf},
we obtain:
\begin{equation}
\tilde{\hat{a}}^{\dagger}=\hat{a}^{\dagger}e^{-i\varphi\num{b}}\hskip0.8cm\tilde{\hat{b}}^{\dagger}=\hat{b}^{\dagger}e^{-i\varphi\num{a}}
\end{equation}
that is:
\begin{equation}
\label{explicitevolveddisp}
\left\{
\begin{aligned}
\tilde{\hat{D}}_{a}(\alpha)&=e^{(\alpha{\hat{a}}^{\dagger}e^{-i\varphi\num{b}}-\alpha^{*}\hat{a}e^{i\varphi\num{b}})}\\
\tilde{\hat{D}}_{b}(\gamma)&=e^{(\gamma{\hat{b}}^{\dagger}e^{-i\varphi\num{a}}-\gamma^{*}\hat{b}e^{i\varphi\num{a}})}.
\end{aligned}
\right.
\end{equation}

Introducing Eq.~(\ref{explicitevolveddisp}) into
Eq.~(\ref{evolvedstate}), we get the time-dependent state of the
two fields:
%\begin{widetext}
%\begin{equation}
\begin{eqnarray}
\label{chissa}
%\begin{split}
\ket{\psi(t)}_{ab} & =&
\tilde{\hat{D}}_{a}(\alpha)\tilde{\hat{D}}_{b}(\gamma)\ket{0}_{a}\otimes\ket{0}_{b}
\nonumber \\
& =&
e^{-\frac{\modul{\alpha}^2}{2}}e^{\alpha{\hat{a}}^{\dagger}e^{-i\varphi\num{b}}}
\ket{0}_{a}\otimes\ket{\gamma}_{b},
%\end{split}
\end{eqnarray}
%\end{widetext}
where the Campbell-Baker-Haussdorff  theorem \cite{MandelWolf} and
the fact that a coherent state is an eigenstate of the
annihilation operator have been used. Using the representation of
coherent states in the Fock number states we have:
%\begin{widetext}
\begin{equation}
\label{chissa1}
%\begin{split}
\ket{\psi(t)}_{ab}=e^{-\frac{\modul{\gamma}^2}{2}}\sum^{\infty}_{n,0}\frac{\gamma^{n}}{\sqrt{n!}}
\ket{\alpha{e}^{-i\varphi{n}}}_{a}\otimes\ket{n}_{b}.
%\end{split}
\end{equation}
%\end{widetext}

If we are experimentally able to set the interaction time  and the
value of the rate of non-linearity so that $\varphi=\pi$ then,
splitting the sum in Eq.~(\ref{chissa1}) into one
over the odd values of $n$ and one over the even ones, the evolved
state of the two initially non-interacting fields takes the form
\cite{SandersMilburn}:
%\begin{equation}
%\label{catstates}
%\begin{split}
%\ket{\psi(\frac{\pi}{\chi})}_{ab}&=e^{-\frac{\modul{\gamma}^2}{2}}\left\{\sum_{n=even}\frac{\gamma^{n}}{\sqrt{n!}}\ket{\alpha}_{a}\right.\\
%&+\left.\sum_{n=odd}\frac{\gamma^{n}}{\sqrt{n!}}\ket{-\alpha}_{a}\right\}\otimes\ket{n}_{b}\Rightarrow\\
%\end{split}
%\end{equation}
\begin{equation}
\label{quasicatstates}
\ket{\psi(\pi/\chi)}_{ab}\propto\ket{\alpha}_{a}\left\{\ket{\gamma}+\ket{-\gamma}\right\}_{b}+\ket{-\alpha}_{a}\left\{\ket{\gamma}-\ket{-\gamma}\right\}_{b}.
\end{equation}

This is a particular expression for an entangled coherent state: it can be reduced to the more familiar form $\ket{\alpha}_{a}\ket{\gamma}_{b}+\ket{-\alpha}_{a}\ket{-\gamma}_{b}$ unitarily acting on the subsystem $b$. To prove the entanglement, we have to show the correlation of the fields of modes $a$ and $b$ as we unitarily transform, gradually, from $\ket{\gamma}_{b}$ to $\ket{\gamma}_{b}\pm\ket{-\gamma}_{b}$. However, this involves another non-linear interaction. We thus discuss an indirect way to prove the production of the entangled coherent state.

In Eq.~(\ref{quasicatstates}), linear superpositions of the coherent states $\ket{\gamma}_{b}$ and $\ket{-\gamma}_{b}$ are
Schr\"odinger cat states:
\begin{equation}
\label{pariedispari}
\begin{split}
\ket{\gamma}_{b}+\ket{-\gamma}_{b}&\propto\sum^{\infty}_{j,0}\frac{\gamma^{2j}}{\sqrt{(2j)!}}\ket{2j}_{b}\\
\ket{\gamma}_{b}-\ket{-\gamma}_{b}&\propto\sum^{\infty}_{j,0}\frac{\gamma^{2j+1}}{\sqrt{(2j+1)!}}\ket{2j+1}_{b}
\end{split}
\end{equation}
which are sometimes called as the even and odd coherent states. As
shown in Eq.~(\ref{quasicatstates}), a time-controlled interaction
of two fields in initially prepared coherent states results in an
entangled state: if we properly normalize state
$\ket{\psi(\pi/\chi)}_{ab}$ and we look to the reduced
density operator of the field $\hat{E}_{a}$ alone (tracing over
$b$ mode), we find an incoherent mixture of two coherent states
which are out of phase by $\pi$ \cite{SandersMilburn}.

Suppose we are, somehow, able to discern the state in which field
$\hat{E}_{a}$ is, after its crossing through our model of highly
non-linear medium. Because of the entangled structure of
Eq.~(\ref{quasicatstates}), the state of field $\hat{E}_{b}$ is
projected onto one of the equally weighted superpositions of
$\pi$-out-of-phase coherent states
$\ket{\gamma}_{b}\pm\ket{-\gamma}_{b}$: it collapses into a
Schr\"odinger cat state. The point, now, is to show how to
reliably discern the state of field $\hat{E}_{a}$.

To achieve this target, we need a $50:50$ beam splitter (BS).  After
passing through a beam splitter, two coherent input fields
$|\alpha\rangle|\beta\rangle$ become \cite{Campos}:
\begin{equation}
\label{bs}
%\begin{split}
\hat{B}_{ac}\ket{\alpha}_{a}\ket{\beta}_{c}=\ket{\frac{\alpha+\beta}
{\sqrt{2}}}_{\tilde{a}}\ket{\frac{-\alpha+\beta}{\sqrt{2}}}_{\tilde{c}},
%\end{split}
\end{equation}
where
$\hat{B}_{ac}\equiv{e}^{\frac{\pi}{4}(\hat{a}^{\dagger}\hat{c}-\hat{a}\hat{c}^{\dagger})}$,
with $a$ ($\tilde{a}$) and $c$ ($\tilde{c}$) as the input (output) modes of the beam splitter.

\begin{figure}[ht]
\centerline{\psfig{figure=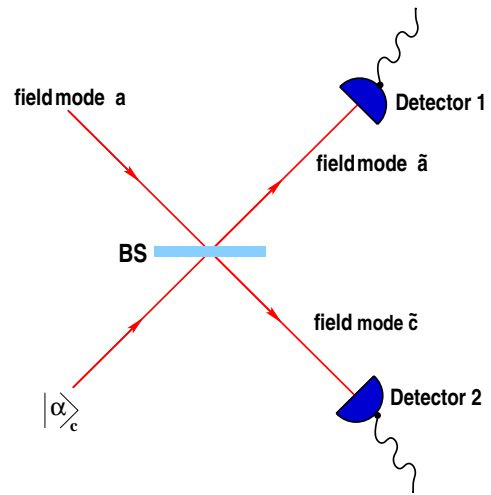,width=6.3cm,height=6.5cm}}
\caption{Detection scheme to infer the state of the field mode
$a$.  In this figure it is shown the symbol used to represent a
photodetector and that for the 50:50 beam splitter (BS). According to
Eq.~(\ref{bs}), Detector $1$ ($2$) clicks just if the state of
mode $a$ is $\ket{\alpha}_{a}$ ($\ket{-\alpha}_{a}$). Using this
read-out scheme, we have the possibility to generate an even or an
odd coherent state of the field mode $b$.} 
\label{detection}
\end{figure}

Thus, taking $\beta=\alpha$ and referring to Fig.~\ref{detection},
we have the following read-out scheme: if the input mode $a$ is in
the state $\ket{\alpha}_{a}$, then Detector $1$ will click,
revealing that some photons arrived at it while Detector $2$ will
not click. In this case, the field mode $b$ will be projected in
the even coherent state. In the opposite occasion, the field mode $b$ 
will be in the odd coherent state. Of course, there
is a possibility to have both the detectors not to click. In this
case, we do not know where the mode $a$ is so we have to repeat
the experiment till we have one detector to click. 

%%%%%%%%%%%%%%%%%%%%%%%%%%%%%%%%%%MISURADEIGATTI%%%%%%%%%%%%%%%%%%%%%%%%%
\section{Detection of a cat state}
\label{misuradeigatti}

The quantum state of an electromagnetic field is completely and unambigously determined once all the moments of its statistical distribution function are determined~\cite{MandelWolf}. To get this task, we need the density matrix $\hat{\rho}$ of the state of the field: it contains all the available information about the given quantum state and, having $\hat{\rho}$, the most complete statistical description of the system can be performed. On the other hand, it is possible to demonstrate a biunique correspondence between the density matrix $\hat{\rho}$ and the Wigner function of a radiation mode~\cite{wignerrho} so that, determining the latter, we are able to fully identify the quantum state of the field. This correspondence 
is useful under an observable point of view: numerous theoretical schemes for the reconstruction of the Wigner function of a radiation mode have been proposed. They are essentially based on heterodyne or homodyne detection~\cite{knightsqueezed}. The mathematical manipulation of the data collected by the homodyne detector allows also for the reconstruction of $\hat\rho$~\cite{vogel,tomography}. Recently, this homodyne tomographic technique has been experimentally realized~\cite{smithey}. This approach, however, may be limited by imperfect detection efficiency.

The procedure described above could certainly be used to determine if our scheme for the non-linear interaction of two weak fields inside a Kerr-enhanced medium has been able to generate a Schr\"odinger cat state. However, as its complete and satisfactory treatment is far beyond our purposes, we want to suggest alternative ways to show the signature of the quantum nature of the state generated. Recently, a sufficient criterion for the inseparability of a continuous variable state has been proposed by Duan {\it et al.}~\cite{duan}. They proved that, for a separable continuous variable state, the total variance of a couple of suitably defined conjugate operators respects a lower bound imposed by the uncertainty relation. For an entangled state, on the other hand, this bound can be violated, providing a sufficient criterion for inseparability of the state. More precisely: assume a continuous variable state composed of subsystems 1 and 2 and take the couple of operators, for both subsystems, $\left\{\hat{\cal X}_{i},\hat{\cal P}_{i}\right\}$ ($i=1,2$) such that $\left[\hat{\cal X}_{i},\hat{\cal P}_{j}\right]=i\delta_{ij}$. Because of this commutation rule, the simultaneous measurement of both the operators is affected by the Heisenberg uncertainty principle. Following~\cite{duan}, we take an arbitrary, non  zero, real parameter $q$ and define the pair of collective operators:
\begin{equation}
\label{variabiliduan}
\left\{
\begin{aligned}
\hat{u}&=\modul{q}\hat{\cal X}_{1}+\frac{1}{q}\hat{\cal X}_{2}\\
\hat{v}&=\modul{q}\hat{\cal P}_{1}-\frac{1}{q}\hat{\cal P}_{2}.
\end{aligned}
\right.
\end{equation}     
For any separable state, it is possible to show that the following inequality holds~\cite{duan}:
\begin{equation}
\label{criterioduan}
S=\valmed{(\Delta{\hat{u}})^2}+\valmed{(\Delta{\hat{v}})^2}\ge{q}^2+q^{-2},
\end{equation}  
with $\valmed{(\Delta{\hat{u}})^2}$ and $\valmed{(\Delta{\hat{v}})^2}$ the average variances, calculated over the state of the system, of $\hat{u}$ and $\hat{v}$ respectively. In what follows, we will often refer to the total variance $S$ as to the separability function.
It is worthwhile to stress that the criterion is just a sufficient condition for inseparability. The established bound could, thus, be exceeded even by 
a continuous variable state that is not separable. We will give an explicit example of this possibility in the following discussion.

The criterion by Duan {\it et al.} leaves a certain freedom in the choice of the couple of operators to use we need to construct $\hat{u}$ and $\hat{v}$. The only limitation imposed is that they have to be conjugate operators.  

In problems related to the investigation of the quantum state of light, it is often useful to consider the phase-space of the quadratures: 
\begin{equation}
\label{quadrature}
\left\{
\begin{aligned}
\hat{x}&=\frac{1}{\sqrt 2}\left(\hat{b}^{\dag}+\hat{b}\right)\\
\hat{p}&=\frac{i}{\sqrt 2}\left(\hat{b}^{\dag}-\hat{b}\right).
\end{aligned}
\right.
\end{equation}
So defined, the quadratures of a field are a couple of conjugate operators, since $[\hat{x},\hat{p}]=i$. Operatively, their probability distribution can be reconstructed by a homodyne detection scheme~\cite{knightsqueezed,YurkeStoler}: the input mode to measure is mixed, at a $50:50$ beam splitter, with the coherent state of a local oscillator whose phase $\vartheta$ is directly controllable during the experiment and whose intensity is so high to consider it classically. The difference between the number of photons in the output modes is, then, measured by two photodetectors (one for each output mode). The data collected are proportional to the expectation value of the operator:
\begin{equation}
\label{omodineoperator}
\hat{O}_{\vartheta}=\frac{1}{\sqrt 2}\left(\hat{b}^{\dag}e^{i\vartheta}+\hat{b}e^{-i\vartheta}\right)
\end{equation}
that, for $\vartheta=0$ coincides with $\hat{x}$, while for $\vartheta=\pi/2$ is equal to $\hat{p}$~\cite{knightsqueezed}. The couple of quadratures 
$\left\{\hat{x}_{a},\hat{p}_{a}\right\}$, $\left\{\hat{x}_{b},\hat{p}_{b}\right\}$ relative to the field modes in Eq.~(\ref{quasicatstates}) are thus well suitable to construct operators $\hat{u}$ and $\hat{v}$ for the joint system of modes $a$ and $b$. What we need, to calculate the total variance function $S$, is just the variance of single quadrature (such as $\valmed{(\Delta\hat{x}_{a,b})^2}$ for example) and the expectation value of correlations as $\valmed{\hat{x}_{a}\hat{x}_{b}}$ or $\valmed{\hat{p}_{a}\hat{p}_{b}}$, measurable collecting the coincidences at the detectors~\cite{kimmunro}.

We take $q=1$, so that the bound value for the separability of the input state is 2. Unfortunately, the calculation of the separability function $S$ for the entangled superposition in Eq.~(\ref{quasicatstates}) leads to $S\ge2$, whatever are the amplitudes $\alpha$ and $\gamma$. As we stressed above, this certainly does not mean that the state is separable but just makes this criterion unsuitable to reveal the entangled nature of the investigated state. Because the behaviour of $S$ is very state-dependent, to bypass the negative result we obtain using Eq.~(\ref{quasicatstates}) directly, we propose the following strategy. Assume we generated, as described at the end of section~\ref{catstates}, an even coherent state of mode $b$. We then mix it with the vacuum of mode $c$ at a $50:50$ BS. Specializing the general rule of a beam splitter in Eq.~(\ref{bs}) for $\beta=0$, the joint state of the BS output modes can be written as:
\begin{equation}
\label{nuovocoerente}
\ket{\phi}_{\tilde{b}\tilde{c}}={\cal N}\left\{\ket{\frac{\gamma}{\sqrt 2}}\ket{-\frac{\gamma}{\sqrt 2}}+\ket{-\frac{\gamma}{\sqrt 2}}\ket{\frac{\gamma}{\sqrt 2}}\right\}_{\tilde{b}\tilde{c}}
\end{equation}
with ${\cal N}=\frac{1}{\sqrt{2(1+e^{-2\gamma^2})}}$. This is an entangled coherent state of two field modes: note that, using this procedure, this output state can be obtained only if the radiation mode $b$ is in an even coherent state. More generally, to describe the state of mode $b$, we can take the density matrix:
\begin{equation}
\label{ro}
\hat{\rho}_{b}={\cal A}\left\{\pro{\gamma}{\gamma}+\pro{-\gamma}{-\gamma}+c(\pro{\gamma}{-\gamma}+\pro{-\gamma}{\gamma})\right\}_{b},
\end{equation}
with $0\le\modul{c}\le1$ and ${\cal A}$ a normalization constant. For $c=0$, the state is a statistical mixture; $c=1$ ($c=-1$) gives us the density matrix of an even (odd) coherent state while, the general case in which $0<c<1$ corresponds to a non optimal generation of the Schr\"odinger cat state. After the action of the $50:50$ BS, the two output modes $\tilde{b}$ and $\tilde{c}$ (whose initial density matrix was $\proj{0}{c}{0}$) will be described by:
\begin{equation}
\begin{aligned}
&\hat{\rho}'_{\tilde{b}\tilde{c}}={\cal A}\left\{\pro{\frac{\gamma}{\sqrt 2},\frac{-\gamma}{\sqrt 2}}{\frac{\gamma}{\sqrt 2},\frac{-\gamma}{\sqrt 2}}+\pro{\frac{-\gamma}{\sqrt 2},\frac{\gamma}{\sqrt 2}}{\frac{-\gamma}{\sqrt 2},\frac{\gamma}{\sqrt 2}}\right.\\
&\left.+c\pro{\frac{-\gamma}{\sqrt 2},\frac{\gamma}{\sqrt 2}}{\frac{\gamma}{\sqrt 2},\frac{-\gamma}{\sqrt 2}}+c\pro{\frac{\gamma}{\sqrt 2},\frac{-\gamma}{\sqrt 2}}{\frac{-\gamma}{\sqrt 2},\frac{\gamma}{\sqrt 2}}\right\}_{\tilde{b}\tilde{c}}.
\end{aligned}
\end{equation}
It is evident that state~(\ref{nuovocoerente}) can be retrieved from this density matrix only if $c=1$ (we have the case of an odd coherent state in input when $c=-1$). 

The quadrature statistics of the two radiation modes are measured by two homodyne detectors, as described above. The collected data allow to estimate the experimental total variance for the input state. This has to be contrasted with what we get calculating the separability function. We thus test, by means of the sufficient criterion for inseparability, the entanglement properties of state~(\ref{nuovocoerente}) in order to infer the state of the radiation mode $b$. Even if, by this means, we are certainly not sure that the state of modes $\tilde{b}$ and $\tilde{c}$ is exactly the entangled coherent state in Eq.~(\ref{nuovocoerente}), we will show the entangled nature of the detected state.

\begin{figure}[ht]
\centerline{\psfig{figure=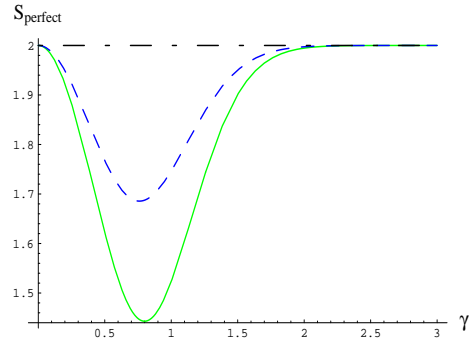,width=6.5cm,height=4.5cm}}
\caption{Plot of $S_{perfect}=\valmed{(\Delta{u})^2}+\valmed{(\Delta{v})^2}$ as a function of 
the coherent state amplitude $\gamma$. The sufficient criterion for inseparability states that, for any separable continuous variable state, this total variance function has to be bound from 
below by a value that, for the calculations shown in this paper, is equal to 2. The effect of different values of the parameter $c$ in the density matrix $\hat{\rho}'_{\tilde{b},\tilde{c}}$ is studied: the dot-dashed curve is for $c=0$, corresponding to to the case of a statistical mixture. The dashed curve is for $c=0.5$ while the solid curve represents the case of a perfectly generated even coherent state of field mode $b$. In this latter case, for the state considered in Eq.~(\ref{nuovocoerente}), the bound is beaten just until $\gamma\simeq2$.}
\label{separabilita}
\end{figure}

If we calculate $S$ for the case of state $\ket{\Phi}_{\tilde{b}\tilde{c}}$, we get:
\begin{equation}
\label{separaperfetta}
S_{perfect}=2\left\{1-\gamma^2\frac{2e^{-2\gamma^2}}{1+e^{-2\gamma^2}}\right\}.
\end{equation}
A plot of $S_{perfect}$ as a function of the amplitude $\gamma$ is given in Fig.~\ref{separabilita}: the total variance function stays below the bound $S=2$ just until $\gamma\simeq2$ (we have $S=1.995$, for $\gamma=2$) and except $\gamma=0$. This gives the signature of the entangled nature of the low intensity state we have generated. The dip of the separability function shows a maximum deviation from the $S=2$ bound equal to $28\%$ for $\gamma\simeq0.8$ and is still about $5\%$ for $\gamma=1.6$. 

Thus, the criterion seems to work sufficiently well for low values of the amplitude of the input coherent states. This is because reducing the value of the amplitude, the entangled coherent state becomes similar to a Gaussian state, that is a class of states for which the total variance criterion for inseparability works very efficiently. In Fig.~(\ref{separabilita}), the separability function~(\ref{separaperfetta}) is compared to what is obtained taking $c=0$ and $c=0.5$ (that is midway between a statistical mixture and a perfectly generated entangled coherent state). The largest effect of an imperfect generation of the input is to reduce the dip of the non-separability well until, for the statistical mixture corresponding to $c=0$, the criterion fails for whatever value of the input amplitude. 
The information we obtain with the detection scheme we are describing give us a sufficient insight into the entangled nature of the state~(\ref{nuovocoerente}).

We now show the robustness of the detection scheme for entanglement with respect to the unavoidable homodyne detector inefficiencies. In what follows we attribute them essentially to losses. Other possible sources of error are the dark-counts at the photodetectors, whose entity depend on the width of the detection window. Their effects can be made negligible taking the detection time short with respect to the characteristic time of a dark-count occurrence. Finally, we neglect the noise introduced by the local oscillators supposing their amplitudes to be well stabilized. 

\begin{figure}[ht]
\centerline{\psfig{figure=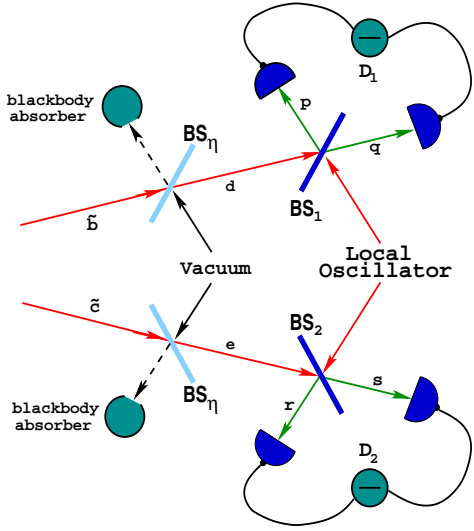,width=6.2cm,height=7.0cm}}
\caption{Sketch of the proposed scheme to infer the value of the total variance for the 
entangled state of modes $\tilde{b}$ and $\tilde{c}$ when the inefficiency of homodyne detectors is
taken into account. With $D_{1}$ and $D_{2}$ we indicate perfect homodyne detectors that have to
measure the quadratures of modes $d$ and $e$. These are the fields transmitted (with
probability $\eta$) by two beam splitters, BS$_{\eta}$, whose effect models the inefficiency of the homodyne detectors. The beams reflected by BS$_{\eta}$ are absorbed by blackbodies. The local oscillators are prepared in very intense coherent states. This allows to treat them as classical fields. Their phases are experimentally controlled and adjustable. BS$_{1}$ and BS$_{1}$ are $50:50$ beam splitters.}
\label{apparatus}
\end{figure}

The effect of loss on the collected data can be modeled replacing a real, inefficient, homodyne detector with a beam splitter (transmittivity $\eta$) followed by a perfect homodyne detector~\cite{YurkeStoler}. In Fig.~\ref{apparatus} the experimental apparatus that considers inefficient homodyne detectors according to the above equivalent model is sketched. Each beam splitter BS$_{\eta}$, that mixes a mode of the signal to measure with a vacuum state, transmits the signal with probability $\eta$ and reflects it with probability $1-\eta$. The amount of reflected input field gives a measure of the losses of the detection apparatus, so that the quantum efficiency of this latter can be identified with $\eta$ itself. To avoid stray light to enter in the perfectly efficient homodyne detector, we can immagine that the reflected output field enters a blackbody absorber. Mathematically, this is equivalent to tracing out the degrees of freedom relative to the absorbed field mode. What the perfect homodyne detectors measures, thus, is the statistic of the quadratures of the transmitted beams, irrespective of the state of the reflected ones. 
\begin{figure}[ht]
\centerline{\psfig{figure=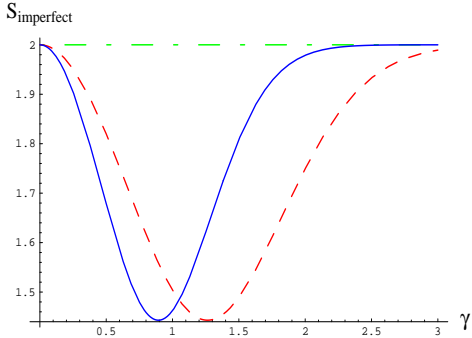,width=6.5cm,height=4.5cm}}
\caption{Behavior of the separability function for state (\ref{nuovocoerente}), as a function of the amplitude $\gamma$, when imperfection in the homodyne detection are taken into account. Here, the dot-dashed curve is for detection efficiency $\eta=0$, the dashed one is for $\eta=0.4$ and the solid curve is for $\eta=0.8$. For lower $\eta$, the minimum values of $S_{imperfect}$ shift toward higher $\gamma$ values: this is because the lower is the efficiency of the homodyne detectors, the more state~(\ref{nuovocoerente}) resembles a Gaussian state.}
\label{condifetto}
\end{figure}

The calculation of the total variance for the quadratures of modes $\tilde{b}$, $\tilde{c}$ when the detectors have an equal quantum efficiency $\eta$ leads to what is shown in Fig.~\ref{condifetto}. The separability function seems to keep its functional features even in the case of imperfect detection. However, some differences, with respect to $S_{perfect}$, appear. As the quantum efficiency of the homodyne detectors reduces, the mimimum of the total variance shifts toward higher values of $\gamma$ (see Fig.~\ref{condifetto}), enlarging the range of amplitudes for which the separability function has values below $2$. 

This effect can be explained directly referring to the probability that the homodyne detector measures a value $x$ of the quadrature $\hat{x}$. Without loss of generality, but avoiding the lengthy calculation relative to state~(\ref{nuovocoerente}), we investigate an even coherent state. We let this state to be mixed with a vacuum field at a BS of transmittivity $\eta$. As before, the resulting transmitted mode is then measured by a perfect homodyne detector while the reflected field is traced out. We finally get the following probability distribution for the in-phase quadrature $\hat{x}$:
\begin{equation}
\label{gaussianhills}
P_{\gamma,\eta,0}(x)=\frac{\left\{2e^{-x^2-2\gamma^2}+e^{-(x-\sqrt{2\eta}\gamma)^2}+e^{-(x+\sqrt{2\eta}\gamma)^2}\right\}}{2\sqrt{\pi}(1+e^{-2\gamma^2})}.
\end{equation} 

It shows that $P_{\gamma,\eta,0}(x)$ is composed of two Gaussian hills centered at $\pm\sqrt{2\eta}\gamma$ plus a third term that, for $\gamma$ sufficiently large, is negligible with respect to the others (even for $\gamma=2$ it is negligible).  The larger is the amplitude $\gamma$, the larger the Gaussian hills are apart~\cite{YurkeStoler}. However, if an efficiency $\eta<100\%$ is considered, the distance between the Gaussian hills for low values of $\gamma$ is strongly reduced and the joint state appears more and more similar to a Gaussian state. To observe well separated Gaussian bells we have to go toward larger $\gamma$. It is possible to show that the same effect is observed in the probability  distribution of an entangled coherent state as Eq.~({\ref{nuovocoerente}}), with the suitable modifications due to the different nature of the state. This explains the behaviour of the separability function for an imperfect homodyne detector.
\begin{figure}
\centerline{\psfig{figure=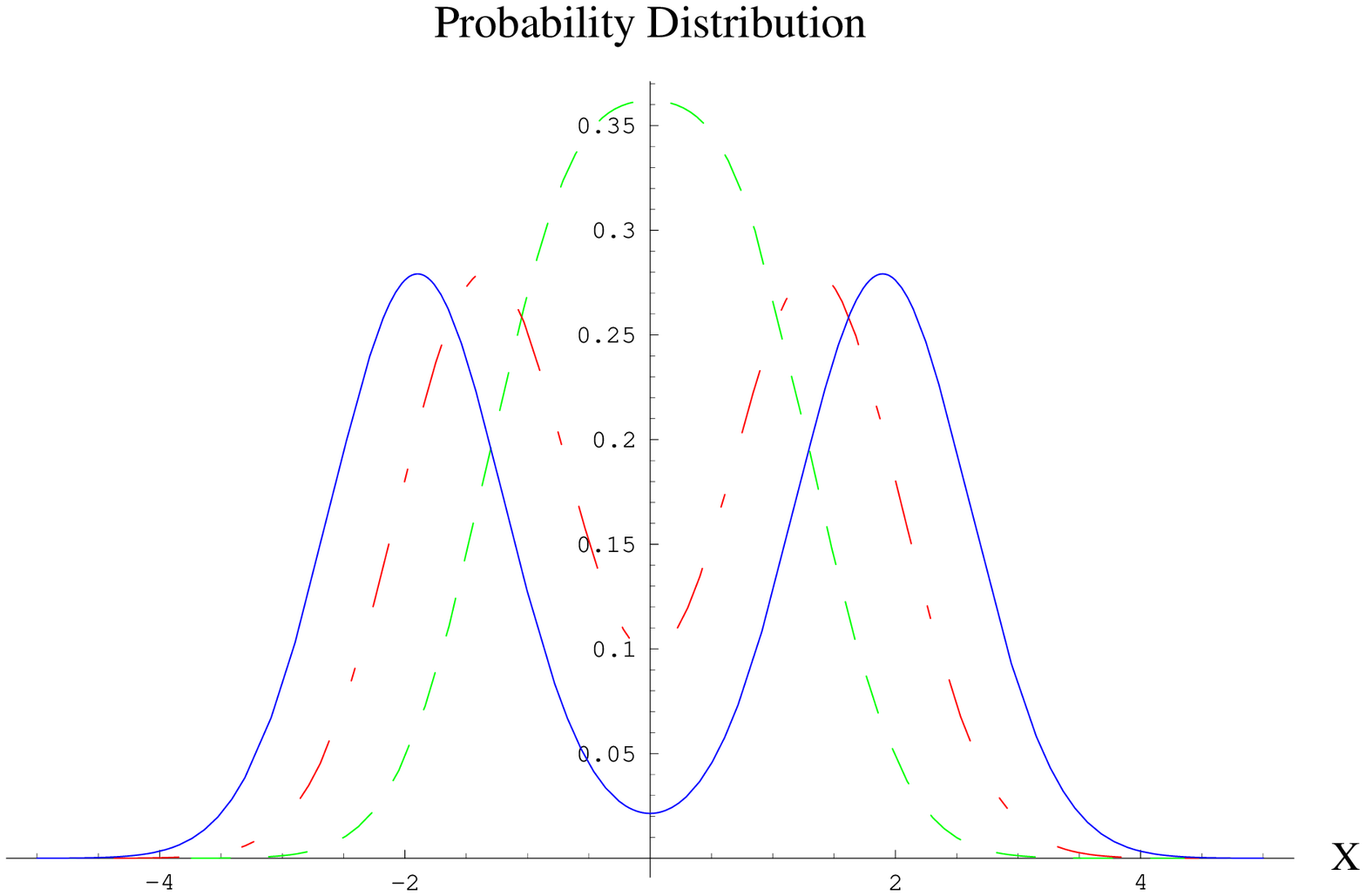,width=6.5cm,height=4.5cm}}
\caption{Plot of $P_{\gamma,\eta}(x)$, for an input field in the even coherent state with $\gamma=1.5$. The effect of different values of $\eta$ is investigated: the dot-dashed curve is for 
$\eta=0.1$, the dashed one is for $\eta=0.4$, the solid one is for $\eta=0.8$.}
\label{colline}
\end{figure} 

However, the success of the total variance criterion for low values of the quantum homodyne inefficiency has not to be considered positively. The imperfections in the detection device tend to hide the features of the input state that appears as a Gaussian state even if, in reality, it is well far to be Gaussian. This means that, in order to reliably test the entangled nature of the input state, a sufficiently large $\eta$ has to be considered. Realistic values of the homodyne detector efficiency range between $0.6$ and $0.85$. In particular, according to our calculations, $\eta=0.8$ seems to be a good trade off between the reliability of the total variance criterion and the realism of an experimentally achievable quantum homodyne efficiency.

From the above analysis, it appears that for $\gamma$ larger than $\gamma\simeq2$, it is not possible to get any information about the entanglement in the input state of modes $\tilde{b}$ and $\tilde{c}$ from the total variance of quadrature amplitudes. We have to look for other detection strategies.      
 
If we still refer to the case of a generated even coherent state in Eq.~(\ref{pariedispari}), the characteristic of the superposition of 
$\ket{\gamma}_{{b}}$ and $\ket{-\gamma}_{{b}}$ is well revealed by homodyne measurements, as pointed out by Yurke and Stoler~\cite{YurkeStoler}. As commented above, the amplitude of the components of an even coherent state can be inferred directly from the probability distribution of the in-phase quadrature. If instead of $\hat{x}$ we measure the statistical distribution for $\vartheta=\pi/2$ in Eq.~(\ref{omodineoperator}), the following function has to be found: 
\begin{equation}
\label{oscillazioni}
P_{\gamma,\eta,\pi/2}(x)=\frac{e^{-x^2}\left\{1+e^{-2(1-\eta)\gamma^2}\cos(2\sqrt{2\eta}\gamma{x})\right\}}{\sqrt{\pi}(1+e^{-2\gamma^2})}.
\end{equation}

\begin{figure}[ht]
\centerline{\psfig{figure=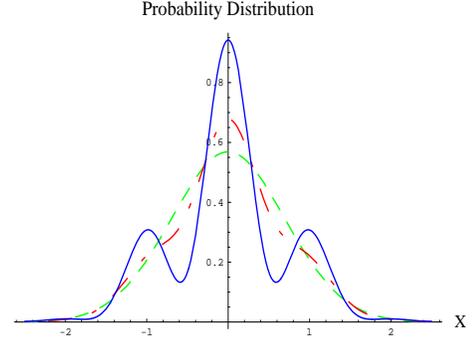,width=6.5cm,height=4.5cm}}
\caption{Probability distribution for a detected homodyne current with $\vartheta=\pi/2$. In this plot, $\gamma=2$ while $\eta$ is scanned: the 
dashed curve is for $\eta=0.4$, the dot-dashed one is for $\eta=0.9$ while the solide curve is for $\eta=0.95$.} 
\label{oscilla}
\end{figure}

This probability oscillates and is moduled, at the same time, by a Gaussian function (solid curve in Fig.~\ref{oscilla}). The oscillations, with a frequency dependent on $\gamma$ as well as on $\eta$, are an evidence of the quantum interferences between the two out-of-phase components of the coherent superposition in the even coherent state. It is, indeed, straightforward to prove that, if instead of a linear combination of $\ket{\gamma}_{b}$ and $\ket{-\gamma}_{b}$ the input state was the mixed state of Eq.~(\ref{ro}) with $c=0$, then this oscillatory behavior would be absent. Thus, a homodyne detection scheme is able to discern a Schr\"odinger cat state from a mixed state of two coherent states which are mutually $\pi$ out of phase~\cite{YurkeStoler}.

It is important to stress the effect, on the oscillatory pattern, due to the Gaussian modulation: as it is evident from Fig.~\ref{oscilla}, even a small reduction of $\eta$ leads, for a small amplitude coherent state, to the disappearance of the oscillations. Mathematically, this is due to the factor $1-\eta$ that governs the width of the Gaussian function in front of the oscillatory term: the smaller $\eta$ is, the more rapidly the Gaussian function goes to zero, washing out the interference pattern. The effect is, clearly, more evident for large values of $\gamma$. This means that this scheme is very vulnerable to the detection inefficiencies.

\begin{figure}[ht]
\centerline{\psfig{figure=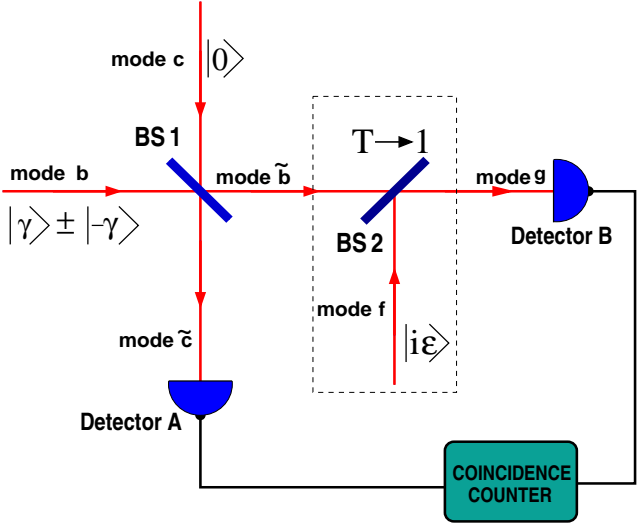,width=8.5cm,height=7.0cm}}
\caption{The figure sketches the apparatus to detect the generated state of mode $b$. As before, the input even (odd) coherent state and the vacuum state of the auxiliary field mode $c$ enter the $50:50$ beam splitter BS$_{1}$ and give rise to an entangled coherent state of the output modes $\tilde{b}$ and $\tilde{c}$. While mode $\tilde{c}$ is directly detected by the photodetector A, mode $\tilde{b}$ is {\it rotated} by the device shown in the dashed box. It consists of the high transmittivity ($T\rightarrow1$) BS$_{2}$ that superimposes a coherent state $\ket{i{\cal E}}$ to the field mode $\tilde{b}$. The transformation thus realized approximates well that of a displacement operator that {\it rotates} the field in $\tilde{b}$ (see Eq.~(\ref{nuovocoerenteruotato})). The trasformed mode is, then, sent to detector B. Both A and B are two {\it On/Off} photodetectors that are able to discriminate the vacuum from any incoming photon. They are not able, however, to reveal the number of incident photons. The coincidences of the counts at the two detectors are measured.}
\label{misuragatti}
\end{figure}

An alternative approach to the investigation of the coherences in the quantum superposition of $\ket{\gamma}_{b}$ and $\ket{-\gamma}_{b}$, when the total variance criterion fails, can be the following. We prepare, once more, the entangled coherent state in Eq.~(\ref{nuovocoerente}) using an even coherent state as input. Similarly to the technique used by Kwiat {\it et al.} in~\cite{zeilinger}, we operate unitarily on one of the components of $\ket{\phi}_{\tilde{b}
\tilde{c}}$. Then, we detect the coincidences between the counts collected by two photodetectors facing the output modes $\tilde{c}$ and $g$ (see Fig.~\ref{misuragatti}).

In the restricted Hilbert space spanned by the orthogonal even and odd coherent states, the rotation of a generic state $\ket{\delta}_{\tilde{b}}=A\ket{\frac{\gamma}{\sqrt 2}}_{\tilde{b}}+B\ket{-\frac{\gamma}{\sqrt 2}}_{\tilde{b}}$ can be performed by a displacement operator $\hat{D}_{\tilde{b}}(i\theta)$ ($\theta\in{\mathbb R}$) that, acting on $\ket{\delta}_{\tilde{b}}$, transforms it into:
\begin{equation}
\ket{\Delta}_{}=Ae^{i\frac{\theta\gamma}{\sqrt 2}}\ket{\frac{\gamma}{\sqrt 2}+i\theta}_{\tilde{b}}+Be^{-i\frac{\theta\gamma}{\sqrt 2}}\ket{-\frac{\gamma}{\sqrt 2}+i\theta}_{\tilde{b}}.
\end{equation}
If we take $\theta\ll\gamma/\sqrt{2}$, on the Bloch-sphere of the restricted Hilbert space that we are treating, the above equation approximates well the state $\hat{R}^{\bf z}_{\tilde{b}}(\alpha)\ket{\delta}_{\tilde{b}}$, with $\hat{R}^{\bf z}_{\tilde{b}}(\alpha)$ the rotation operator, by an angle $\alpha/2={\sqrt 2}\gamma\theta$, around the Bloch-sphere {\bf z}-axis. Experimentally, such a rotation can be accomplished superimposing mode $\tilde{b}$ on the coherent state $\ket{i{\cal E}}$ of an auxiliary mode $f$, using a beam splitter with a high-transmission coefficient $T$. As shown in~\cite{kimrotation}, the transformation operated by the $T\rightarrow1$ beam splitter (BS$_{2}$ in Fig.~\ref{misuragatti}) gives an entangled state of two output field modes. The reduced density matrix that describes the state of the output mode $g$ only can then be approximated by that of a displaced state according to $\hat{\rho}_{g}=\hat{D}_{\tilde{b}}(i{\cal E}\sqrt{1-T})\hat{\rho}_{\tilde{b}}\hat{D}^{\dag}_{\tilde{b}}(i{\cal E}\sqrt{1-T})$ where $\hat{\rho}_{\tilde{b}}$ is the density matrix of the input field~\cite{kimrotation}.

If we vary ${\cal E}$ in such a way that $\theta={\cal E}\sqrt{1-T}\ll\gamma/\sqrt{2}$, that is: 
\begin{equation}
\label{rotazione}
{\cal E}=\frac{\alpha}{2\sqrt{2(1-T)}\gamma},  
\end{equation}
then, just controlling the amplitude of this ancillary mode we can appropriately {\it rotate} the state of mode $\tilde{b}$ by $\alpha/2$. 

Following this lines, the entangled coherent state in Eq.~(\ref{nuovocoerente}), after the interaction of $\tilde{b}$ with $f$ at the {\it rotation beam splitter} BS$_{2}$, becomes:
\begin{equation}
\label{nuovocoerenteruotato}
\ket{\Phi}_{g\tilde{c}}={\cal N}\left\{e^{i\frac{\alpha}{2}}\ket{\frac{\gamma}{\sqrt 2}}\ket{\frac{-\gamma}{\sqrt 2}}+e^{-i\frac{\alpha}{2}}\ket{\frac{-\gamma}{\sqrt 2}}\ket{\frac{\gamma}{\sqrt 2}}\right\}_{g\tilde{c}}.
\end{equation}

The output modes $\tilde{c}$ and $g$ are then sent to detector A and B respectively, where the photo-count coincidences are revealed. We model the photodetectors as {\it Geiger-like On/Off Photo-Detectors} (\oodet). These particular devices just discriminate the vacuum from an input with any photons, irrespective of what the photon number is \cite{parisbonifacio}. Again, losses make the quantum efficiency of the an {\oodet} not optimal, so that some of the coincidences will be missed. Quantum mechanically, an inefficient {\oodet} can be described by means of a suitably defined {\it Positive-Operator-Valued-Measure} (POVM), that is a set of diagonalizable, having positive-eigenvalues, projection operators. The appropriate choice for this case is the following: 
\begin{equation}
\label{povm}
\left\{
\begin{aligned}
&\Pi^{(i)}_{\stackrel{no}{click}}({\eta})=\sum_{n,0}^{\infty}(1-\eta)^{2n}\proj{n}{i}{n}\\
&\Pi^{(i)}_{click}({\eta})=\one-\Pi^{(i)}_{\stackrel{no}{click}}({\eta}),
\end{aligned}
\hskip0.5cm(i={\text {A, B}})\right.
\end{equation}   

We assume the same efficiency $\eta$ for both the detectors. In the above definitions $1-\eta$ is the probability that a single detector lacks to reveal a photon and the sum over all the photon populations takes in account the impossibility to distinguish the photon number in the incident field.
 Within the POVM formalism, the probability that the {\oodet} click within the same detection window is given by:
\begin{equation}
\label{doppioclick}
\begin{aligned}
&P_{2clicks}(\gamma,\eta,\alpha)={\it Tr}_{g\tilde{c}}\left\{\proj{\Phi}{g\tilde{c}}{\Phi}\Pi^{(\text A)}_{click}({\eta})\Pi^{(\text B)}_{click}({\eta})\right\}=\\
&{\cal C}\left\{\left(e^{\frac{\gamma^2}{2}}-e^{\frac{(1-\eta)\gamma^2}{2}}\right)^{2}+\cos{\alpha}\left(e^{-\frac{\gamma^2}{2}}-e^{-\frac{(1-\eta)\gamma^2}{2}}\right)^{2}\right\}, 
\end{aligned}
\end{equation} 
with ${\cal C}=2e^{-\gamma^2}{\cal N}^{2}$.
Let us look to the asymptotic behaviour of the above probability: for $\eta\rightarrow{0}$, that is for highly inefficient detectors, we get $P_{2clicks}(\gamma,0,\alpha)\rightarrow0$. If the limit $\eta\rightarrow{1}$ is instead considered, Eq.~(\ref{doppioclick}) shows that $P_{2clicks}(\gamma,1,\alpha)\rightarrow2\left(1+e^{-\gamma^2}\cos\alpha\right)$. The oscillating part in $P_{2clicks}(\gamma,1,\alpha)$ is the evidence of the coherences established in the even coherent state. Unfortunately, it has an exponential pre-factor that can cause experimental difficulties (for $\gamma\simeq2$ and $\eta\simeq0.75$, a visibility of about $1\%$ is achieved). Furthermore, possible errors connected to the {\it rotation} of a Schr\"odinger cat state have to be considered.

Once the entangled nature of the input state $\ket{\Phi}_{\tilde{b}\tilde{c}}$ is recognized, however, the inference about the state of field mode $b$, generated by the non-linear interaction via double-EIT, proceeds along the lines outlined before.
 
From the above discussion, it is clear that the main problems related to the described schemes rely on the quantum efficiency of the detection devices. As we have seen, imperfections in the homodyne detectors, for example, are responsible for the disappearance of the oscillation pattern in the probability distribution relative to the out-of-phase quadrature. For the case of a Schr\"odinger cat state, this results in a loss of information relative to the quantum superposition of two coherent states. It is worthwhile to stress that for high value of $\eta$, the oscillations in $P_{\gamma,\eta,\pi/2}(x)$ survive even for large values of $\gamma$.

The hard task represented by the detection of an even (odd) coherent state or, more generally, of an entangled coherent state can be made easier by the recently improved sensitivity of the available detectors. Values of $\eta\sim0.8$ are realistic for a detector operating in a Geiger-mode and very sensitive photo-detectors, with an efficiency as high as $\simeq0.93$, have been developed. They however require a low temperature to properly operate~\cite{takeuchi}.

%%%%%%%%%%%%%%%%%%%%%%%%%%%%%%%%%%CONCLUSIONS%%%%%%%%%%%%%%%%%%%%%%%%%%%%%

\section{Conclusions}
\label{conclusions}
We have proposed a fully quantized picture of the model for
double-EIT recently discussed by Petrosyan and Kurizki
\cite{PetrosyanKurizki}. Our approach is based on a full
Hamiltonian method. This allows to bypass the analytical solution
of the equations of motion of the  atomic density matrix elements. It
simplifies the computational problems related to systems that
involve many atomic energy levels coupled by electromagnetic
fields.

As shown for the first time in \cite{LukinImamoglu}, the
non-linear interaction of two beams of light that pass through a
dense atomic medium is optimized by a double-EIT regime in which
both the fields propagate with a strongly reduced, equal, group
velocity. The atomic model proposed in \cite{PetrosyanKurizki}
seems to be a good candidate to the experimental realization of
such a physical condition. In our fully quantum version of this
model, we have shown that the quantum dynamics of one of the
interacting beams is dramatically dependent on the intensity of
the second and viceversa. Our results are consistent with those
reported in \cite{LukinImamoglu}. We suggested the Pr:YSO~\cite{turukhin} 
crystal as a possible candidate to physically embody the discussed atomic model.

Starting from these results, we have written an  effective
interaction Hamiltonian that, when the field is initially prepared
in two coherent states, leads to the possibility to
generate entangled coherent states and even or odd coherent states (Schr\"odinger cat states). In order to investigate the quantum features of a generated even (odd) coherent state, 
we described a scheme able to measure the total variance function of the involved modes quadratures~\cite{kimmunro}. For low amplitudes of 
the investigated entangled coherent state, the value of this total variance is well below the bound imposed by the 
inseparability criterion suggested in~\cite{duan}. We also investigated the effects of an 
imperfect non-linear process and of detection inefficiencies on the above results. We found our scheme robust against homodyne 
detection losses. For the cases in which the total variance criterion is unuseful, we suggested a detection scheme based on the unitary manipulation of one of the modes in the entangled coherent 
state followed by the detection of coincidences of photo-counts. We discussed possible sources of errors and experimental difficulties we expect 
to appear in a real experiment, finding that an optimal detection protocol basically requires high efficiency of the detectors. 

%%%%%%%%%%%%%%%%%%RINGRAZIAMENTI E FINANZIAMENTO%%%%%%%%%%%%%%%%%%%%%%%%%%%%%%%

\section*{Acknowledgments}
We want to thanks Dr. Anatoly Zayats for the useful discussions about optical Kerr effect enhancement.
This work is supported by the UK Engineering and Physical Science Research Council through GR/R33304. B.H. acknowledges the financial support from the  Korean
Ministry of Science and Technology through the Creative Research Initiative
Program. M.P. acknowledges IRCEP (International Research Centre for Experimental Physics, Queen's University of Belfast) for supporting his studentship.
%%%%%%%%%%%%%%%%%%%%%%%%%%%%%%%%%BIBLIOGRAFIA%%%%%%%%%%%%%%%%%%%%%%%%%%%%%%%%%%%%%%%%%%%%%%%%%%%%%%%%%%%%%%%%%%%%%%%%%%%%%

%\end{multicols}

\end{document}